\documentstyle[twocolumn,aps,prl,epsf,floats]{revtex}
\begin{document}
\draft
\flushbottom
\twocolumn[\hsize\textwidth\columnwidth\hsize\csname @twocolumnfalse\endcsname 
\preprint{} 
\title{Universal scaling, beta function, and metal-insulator 
transitions }
\author{D.N. Sheng and  Z.Y. Weng} 
\address{Texas Center for Superconductivity, University of Houston, Houston,
 TX 77204-5506 }  
\tightenlines
\widetext
\maketitle 
\date{today}
\begin{abstract} 

We demonstrate a {\it universal} scaling form of longitudinal resistance in
the quantum critical region  of metal-insulator transitions, based on 
numerical results of three-dimensional Anderson transitions (with and 
without magnetic field), two-dimensional quantum Hall plateau to 
insulator transition, as well as experimental data of the recently 
discovered two-dimensional metal-insulator transition. The associated
reflection symmetry and a peculiar {\it logarithmic} form of the beta
function exist over a wide range in which the resistance can change by more
than one order of magnitude. Interesting implications for the two-dimensional 
metal-insulator transition are discussed.

\end{abstract}
]

\narrowtext

\newlength{\mysize}
\def\loadepsfig#1{
 \def\figname{#1}
 \vbox to 10pt {\ }
 \vbox{ \hbox to \hsize {
   \mysize\hsize    \advance \mysize by -38pt
   \def\epsfsize##1##2{\ifdim##1>\mysize\mysize\else##1\fi}
   \hfill \epsffile{\figname.eps} \hfill
        } }
 \vbox to 10pt {\ }
 }
\def\gloscale{0.8}
\def\dispfig#1{
  \def\figname{#1}
  \def\epsfsize##1##2{\gloscale##1}
  \settowidth{\mysize}{ \epsffile{\figname.eps} }
  \parbox{\mysize}{ \epsffile{\figname.eps} }
  }
\newlength \mygapsize

The scaling theory\cite{abhm} predicted that the noninteracting electrons are 
always localized in two-dimensional (2D) disordered systems. Recently, a new 
scaling argument\cite{the1} was put forward in order to accommodate the
newly-found 2D metal-insulator transition (MIT)\cite{krav} in 
zero magnetic field ($B=0$), where the Coulomb interaction presumably becomes
very important\cite{krav}. Although the microscopic mechanism remains 
unclear\cite {the2,the3,dns}, without violating any general scaling
principles the authors {\it assumed} the following leading behavior of the 
``beta function'' $\beta(g)=d[ln(g)]/d[ln(L)]$ for large conductance $g$ at a 
finite length scale $L$:
\begin{equation}\label{1}
\beta(g)=(d-2)+ A/g^{\alpha}+ ...
\end{equation}
in which $A$ becomes {\it positive} in the aforementioned $B=0$ MIT 
systems\cite{krav,exper1,exper2}, leading to a metallic phase ($\beta>0$) 
at the dimensionality $d=2$.  

Since $\beta(g)<0$ at small $g$ (localized region), the beta function 
is then no longer a monotonic function and has to change sign at some finite
$g=g_c$, which corresponds to a quantum critical point. Experimental 
measurements have indicated\cite{krav} an exponential form for the conductance
with a peculiar reflection symmetry relating the conductance and the resistance 
on both sides of MIT, which implies\cite{the1} the following
{\it logarithmic} form of the beta function in the quantum critical 
region (QCR):   
\begin{equation}\label{2}
\beta(g)= \frac{1}{\nu}ln (g/g_c).
\end{equation} 
In particular, $\nu$ here is the correlation length exponent, and 
(\ref{2}) holds at a wide range ($1/4<g/g_c<4$) far beyond a simple small 
variable expansion around $g=g_c$. 

The logarithmic form of the beta function (\ref{2}) looks quite remarkable. 
Recall that in strong localization limit one may find, exactly, 
$\beta(g)=ln(g)- constant $. But the inverse exponent $1/\nu$ does not 
show up in front of $ln(g)$ as in (\ref{2}) and the corresponding 
behavior of $g$ should be quite different. So far there still lacks a
good theoretical understanding of (\ref{2}) from a microscopic model. 
Nevertheless, one may ask an equally important question: whether 
(\ref{2}) is a property of the beta function unique for the $B=0$ 2D MIT 
system or it actually represents a generic scaling behavior of  quantum 
phase transitions including other MIT systems with different symmetry and 
dimensionality. Unfortunately, so far there is no direct 
anwser to this question as  how the {\it scaling function} of
conductance behaves and what is the form of the beta function in the 
QCR of various MIT systems are not known, although a lot of
efforts have been focused on the critical conductance and
exponent within each universality class\cite{class,three,qhe}.

In this paper, we present direct numerical evidence showing that the beta
function (\ref{2}) in fact holds for the following systems as well:
three-dimensional (3D) Anderson transitions with and without magnetic field 
(representing orthogonal and unitary classes, respectively); the 2D electrons
in strong magnetic filed, i.e. the quantum Hall effect (QHE) system.
Strikingly, $\nu \beta(g)=ln(g/g_c)$ is found to be a {\it universal} 
function in the QCR where $g/g_c$ may change up to two orders of magnitude. 
Correspondingly the resistance is of an exponential form 
$\rho_{xx}\propto e^{-s}$ with $s=\pm (c_0 L/\xi)^{1/\nu}$ which also implies
a reflection symmetry in the same region (here $\xi$ is correlation length,
and $c_0$ is a non-universal dimensionless constant $\sim O(1)$). Thus (2) may 
well represent a ``super'' universality property associated with general 
quantum 
phase transitions.  Furthermore, deep into the metallic region, the beta 
function shows distinct behavior depending on how the resistance $\rho_{xx}$ 
deviates from the exponential form: 3D MITs and the $B=0$ 2D MIT 
experimental data seem to belong to the same group 
where $d[ln(1/\rho_{xx})]/ds $ is a monotonically {\it decreasing} function of
scaling variable $s$ in the whole scaling region; on the other hand, the QHE 
system falls into a different group where $d[ln(1/\rho_{xx})]/ds $ becomes 
a monotonically {\it increasing} function of $s$. Interestingly, the 
experimental data of superconductor-insulator transition\cite{super}
also fall into the second group, in accord with the speculation\cite{fisher}
that the MIT in the QHE and superconductor-insulator transition may belong
to similar universality class. 

We consider disordered electron systems described by the Anderson 
Hamiltonian \cite{anderson}:
\begin{eqnarray*} 
H=-\sum_{<ij> } e^{i a_{ij}}c_i^+c_j + H.c. +\sum _i w_i c^+_i c_i , \nonumber
\end{eqnarray*} 
where the hopping integral is taken as the unit, and $c_i^+$ is a  
fermionic creation operator with $<ij>$ referring to the nearest neighboring 
sites.  A uniform  magnetic flux per plaquette (along {\bf $z$} direction) can 
be imposed by requiring   $\phi=\sum _ {\Box} a_{ij}=2\pi/M$,
where the  summation runs over four links around a plaquette in the {\bf $x-y$} plane.  $w_i$ is  a random potential  uniformly distributed between 
(-$W/2,W/2$).

We first study the 3D electron system without magnetic field 
($a_{ij}=0$)  which belongs to the orthogonal class. 
The longitudinal conductance $G_{xx}$ is calculated using Landauer
formula\cite{lan}.  By changing the disorder strength $W$, a 
metal-insulator transition is found at a critical disorder strength 
$W_c=16.5$\cite{three} at the Fermi energy $E_f=0$, with a critical 
conductance $G_c=0.37 $ (in units of $e^2/h$) and correlation exponent
$\nu=1.6$. All the data at different sample sizes ($L=8,10,12, 14$ and $16$) 
can be then collapsed onto two branches as a function of $L/\xi$ as shown in
Fig. 1(a) ($\diamond$ curve). Note that the resistance data are plotted in 
the figure. (More than $2000$ configurations are taken in the average for 
$L=16$, and more for smaller $L$'s.) A 3D MIT is similarly obtained in the 
presence of strong magnetic field (unitary class). We have chosen two 
different flux strengths $\phi=2\pi/M$: $M=5$ at sample sizes $L=10$ and 
$15$; and $M=4$ at sample sizes $L=8,12,$ and $16$, respectively.  All the 
longitudinal resistance data with different $\phi$'s and $L$'s again can be
scaled onto two branches ($+$ curve in Fig. 1(a)). At the critical point, 
$W_c=18.3$,  $\nu=1.43$ \cite {three} and  $G_c=0.294$ at $E_f=0$. 
\begin{figure}[ht!]
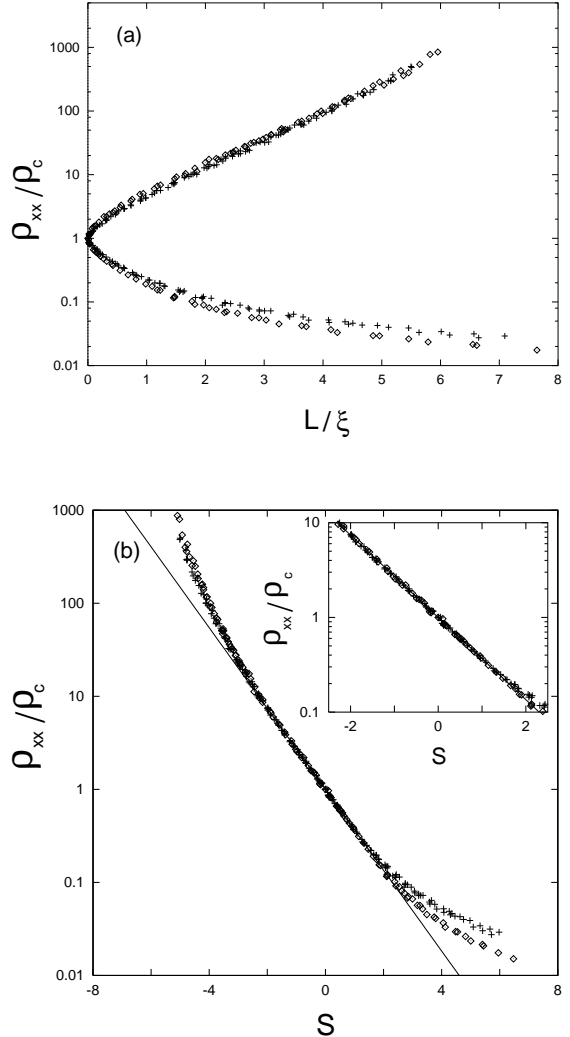
 % fig 1
\loadepsfig{mitfig1a}
\loadepsfig{mitfig1b}
\caption[\ ]{ (a) Finite-size resistance ratio $\rho_{xx}/\rho_c$ as a 
scaling function of $L/\xi$ for 3D Anderson transitions
in the absence ($\diamond$) and presence ($+$) of magnetic field.
(b) $\rho_{xx}/\rho_c$ as a function of the scaling variable $s=\pm
(c_0L/\xi)^{1/\nu}$. The insert: the enlarged quantum critical region where
$\rho_{xx}/\rho_c=e^{-s}$.}
\label{fig1}
\end{figure}

Since the universality of the MIT in the unitary class is distinct from the one
of the orthogonal class as expected in the scaling theory\cite{class,three},
two scaling curves shown in Fig. 1(a) are generally different from each 
other. However, if we re-plot the data in terms of the scaling variable
\begin{equation}\label{3}
s=\pm  \left(c_0L/\xi\right)^{1/\nu},
\end{equation}
where the sign $+(-) $ corresponds to the metallic (insulating) branch, 
two curves of longitudinal resistance in the QCR can be precisely scaled 
together as shown in Fig. 1(b). Here the dimensionless constant $c_0$ has the
non-universal values $2.27$ and $1.82$ for orthogonal and unitary class, 
respectively. As shown in the insert of Fig. 1(b), the resistance in the 
transition region well follows a simple exponential form 
\begin{equation}\label{4}
\rho_{xx}/\rho_c=exp(-s)
\end{equation}
over a rather broad region: $-2<s<2$ or $1/8<\rho_{xx}/\rho_c<8$.
Because of such a wide range of $s$ (instead of a small parameter
expansion), the exponential behavior appears very robust. In the same 
region, one always finds the so-called reflection symmetry: 
$\rho_{xx}(s)/\rho_c =\rho_c/\rho_{xx}(-s)$ between the metallic and 
insulating branches. 

Now let us consider a qualitatively different MIT in the QHE system where 2D
electron gas is subjected to a strong magnetic field. By tuning the Fermi 
energy (or the density of electrons) near the lowest Landau Level (LL), an 
insulator to metal transition can be induced which is characterized by 
a one-parameter scaling theory\cite{qhe} with an exponent $\nu=7/3$ and 
$\rho_{c}=1$ (in units of $h/e^2$)\cite{dns1}. The Hall conductance here is
calculated using Kubo formula. By going to large sample sizes ($L=24,32,48,56,
64$), we were able to obtain the scaling behavior of $\rho_{xx}$ for the QHE
systems.  In Fig. 2, $\rho_{xx}$ is plotted as a function of scaling variable
\begin{figure}[hb!]
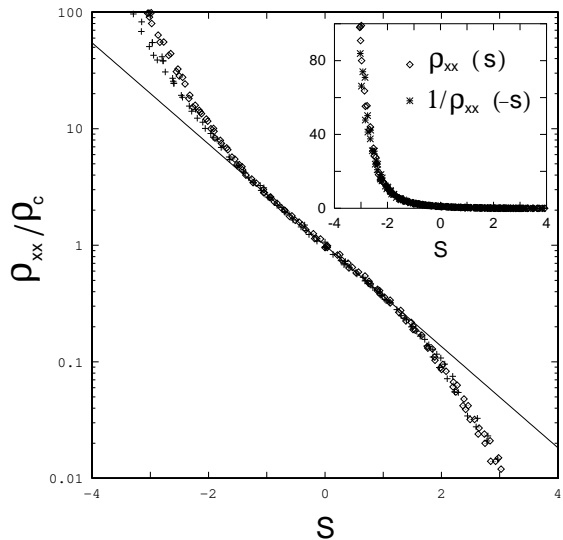
 % fig 2
\loadepsfig{mitfig2}
\caption[\ ]{$\rho_{xx}/\rho_c$ as a
function of the
scaling variable $s$ for the QHE systems both at weak ($\diamond$, $W=1$)
and strong ($+$, $W=4$) disorder strengths.
The insert: the reflection symmetry of $\rho_{xx}$ at $W=1$.       } 
\label{fig2}
\end{figure}
$s$ defined in (\ref{3}). Again the resistance exhibits the same exponential 
dependence $\rho_{xx}=exp(-s)$ (solid line) in the critical
region covering  a similar wide range of resistance ($1/5<\rho_{xx}/\rho_c<5$)
as in the 3D MITs. In Fig. 2, two different disorder 
strengths, $W=1$ and $W=4$, are considered which represent weak and strong LL 
coupling limit, respectively. The corresponding scaling functions start to
deviate from the exponential form beyond the critical region and 
simultaneously become $W$-dependent in the insulating region.

We would like to point out an interesting reflection symmetry for the $W=1$ 
case: as shown in the insert of Fig. 2, $\rho_{xx}(s)$ on the insulating side
and $1/\rho_{xx}(-s)$ on the metallic side perfectly coincide with each other
over the whole scaling  region  and covering a resistance range 
$1/100<\rho_{xx}<100$ which  is way beyond the critical region. Our 
interpretation is that at weak disorder ($W=1$), the 
particle-hole symmetry is still approximately maintained near the
lowest LL such that the Hamiltonian is self-dual\cite{fisher,shahar} in the 
Chern-Simon boson language, which then leads to the wide range of the
reflection symmetry. By contrast, when disorder is strong and all the LLs 
are coupled together without the particle-hole symmetry, the reflection
 symmetry only exists around the QCR where the exponential behavior 
(\ref{4}) is followed.

As demonstrated by the above numerical calculations, the scaling function
of longitudinal resistance shows a universal exponential behavior over a wide
range in 3D and QHE MITs. In Fig. 3, these data are plotted together with the 
experimental data obtained in the $B=0$ 2D MIT in the $Si$ 
sample\cite{krav}. Note that the experimental data were measured at finite
temperature so the length scale $L$
should be replaced by the dephasing length $L_{in} \propto T^{-1/z}$.
(Here $z=1$ is the dynamical exponent).  The correlation length 
$\xi \propto 1/T_0 \propto |\delta _n|^{-\nu}$ in the transition 
region\cite{krav} ($\delta_n $ is the electron density
measured from the critical point). So scaling variable $s$ in this case 
becomes $\pm (c_0 T_0/T)^{1/\nu}$ with $\nu=1.6$ and $c_0$ is a dimensionless
constant.  In Fig. 3, a universal scaling function of the longitudinal 
resistance is clearly shown  
for all these systems in the QCR (with $-1.5<s<1.5$) despite their different
symmetry classes, dimensionalities, and microscopic mechanisms of the MIT.
\begin{figure}[hbt!]
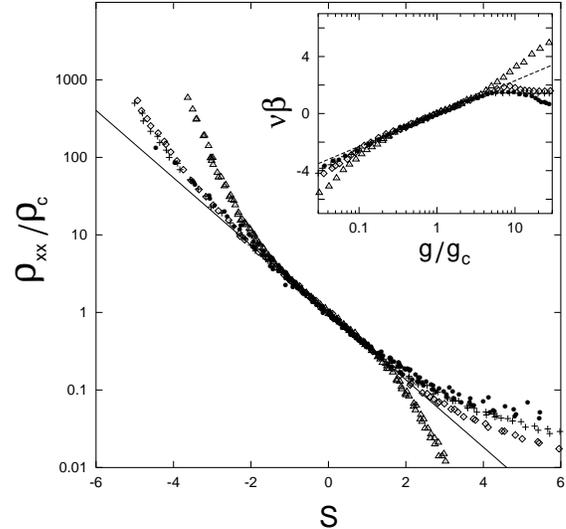
 % fig 3
\loadepsfig{mitfig3}
\caption[\ ]{Experimental data\cite{krav} of $B=0$ MIT ($\bullet$) are
plotted together with the resistances calculated for the 
systems of 3D Anderson transition ($\diamond$ and $+$) and the QHE ( 
$\triangle$). The
corresponding $\beta$ functions are shown in the insert 
which precisely
follow the logarithmic form (\ref{2}) in the QCR.}
\label{fig3}
\end{figure}
The corresponding beta functions for those 
systems are shown in the insert
of Fig. 3 if we define $g\equiv 
1/\rho_{xx}$, where a straight dashed
line represents the logarithmic form of 
(\ref{2}) which can be obtained
straightforwardly from (\ref{4}). Note 
that the beta function is
multiplied by the critical exponent $\nu$ 
in the insert such that the
resulting function becomes {\it universal} 
in the QCR.
                                                                                           
Furthermore, we would like to comment on an interesting trend in the metallic
region for those systems. In weak disorder limit of 3D MITs, the resistance 
approaches to zero in power law:  $\rho_{xx}\sim (\xi/L)=c_0 s^{-\nu}$.
The curve for $B=0$ 2D MIT system follows very closely to the ones
of 3D MITs on the same side of the solid line in Fig. 3 as it 
drops to zero slower than $exp(-s)$.
In contrast, in the QHE system, the resistance deviates the solid line on the 
opposite side which means it approaches to zero even quicker than in the QCR.
This behavior can be easily seen in its asymptotic form: 
$\rho_{xx} \sim \sigma_{xx}\sim exp(-s^{\nu}/c_0)$ at large $s$ limit 
(since in the QHE plateau region electrons are also localized such that at 
large $L/\xi$ limit  $\sigma_{xx}\sim exp(-L/\xi)$,  $\sigma_{xy}=1$ and 
$\rho_{xx} \sim \sigma_{xx}$). One may then define a generalized 
dimensionless function as follows: 
\begin{equation}\label{5}
\beta_1\equiv d[ln(g)]/ds=\nu\beta/s
\end{equation}
As shown in Fig. 4, all the data fall onto the straight line with $\beta_1=1$
in the QCR. In the metallic phase the distinctive large-$s$ behavior of 
$\beta_1$ separates the
metallic regime into two regions. METAL denotes the region where $\beta_1$ 
scales to zero which is followed by the 3D MITs as well as the experimental
data of the $B=0$ 2D MIT system. On the other hand, $\beta_1$ for the MIT
in the QHE system diverges to infinity at $s\rightarrow \infty$, which is 
denoted as METAL(B) region known as the ``Bose'' metal following the 
theoretical description\cite{fisher}.
$\beta_1$ for disorder-tuned superconductor-insulator transition\cite{super} in
the metallic region is also plotted in Fig. 4 ($*$ curve) which indeed shows
a quick increase like in the QHE system. (Here the scaling variable $s$ is
of the same form used in plotting the experimental data of B=0 2D MIT.) 
According to Ref.\onlinecite{fisher},
these two systems should belong to the same category as classified by the
``Bose'' metal here.      
\begin{figure}[hbt!]
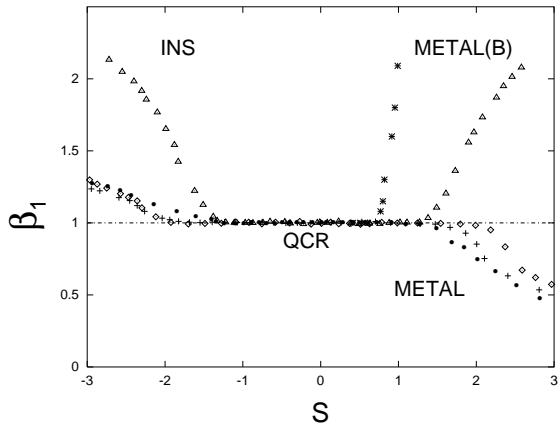
 % fig 4
%\centerline{\epsfig{file=mitfig1a.eps,height=3in,width=3in}}
%\centerline{\epsfig{file=mitfig1b.eps,height=3in,width=3in}}
%\vspace{10pt}
\loadepsfig{mitfig4}
\caption[\ ]{ The function $\beta_1$ defined in (\ref{5}) as a
function of the scaling variable $s$ for those systems shown in Fig. 3. In
addition, the data for a disorder-tuned superconductor-insulator
transition\cite{super} in the metallic regime are also shown for 
comparison ($*$).           } \label{fig4}
\end{figure}       

We conclude by making several comments on the nature of the $B=0$ 2D MIT 
systems based on the present work.  First, no matter what the microscopic 
mechanism is, such a 2D MIT seems to belong to a quantum phase transition 
instead of a classical phase transition (or crossing over): The experimental 
data of the resistance precisely coincides with those of other known MITs in
the QCR, plotted as a function of the scaling variable 
$(L_{in}/\xi)^{1/\nu}$, 
which covers a range of the resistance by more than one order of magnitude. 
Second, the reflection symmetry of resistance\cite{krav} is the natural 
consequence of the
universal resistance scaling in the QCR. Finally, the metallic phase behaves
more like a normal metal than a superconductor as revealed by the 
classification based on the $\beta_1$ function shown in Fig. 4.

{\bf Acknowledgments} -
The authors would like to thank S.V. Kravchenko for stimulating discussions
and providing his experimental data for comparison. 
This work is supported by the ARP grant No. 3652707, and by the State 
of Texas through Texas Center for Superconductivity at University of Houston.


\begin{references} 
\bibitem{abhm} E. Abrahams, P. W. Anderson, D. C. Licciardello, and 
T. V. Ramakrishnan,  Phys. Rev. Lett.  {\bf 42}, 673 (1979).
\bibitem{the1} V. Dobrosavljevic, E. Abrahams, E. Miranda, and S. Chakravarty, 
Phys. Rev. Lett.  {\bf 79}, 455 (1997).
\bibitem{krav} S. V. Kravchenko et al.,   Phys. Rev. B {\bf 50},
8039 (1994); {\it ibid.} {\bf 51}, 7038 (1995);
S. V. Kravchenko et al., Phys. Rev. Lett. {\bf 77}, 4938 (1996).
D. Simonian  et al.,
 Phys. Rev. B {\bf 55},  R13421 (1997). 
\bibitem{the2} S. Chakravarty, L. Lin, and E. Abrahams, Phys. Rev. B
{\bf 58}, R559 (1998); S. Chakravarty et al., 
 cond-mat/9805383 (1998); 
C. Castellani et al.,    Phys. Rev. B {\bf 57}, R9381 (1998);
 Q. Si and C. M. Varma, Phys. Rev. Lett. {\bf 81}, 4951 (1998).
\bibitem{the3} D. Belitz and T. R. Kirkpatrick, Phys. Rev. B {\bf 58}, 8214 
(1998);  P. Phillips et al., Nature {\bf 395}, 253 (1998).
\bibitem{dns} D. N. Sheng and Z. Y. Weng, cond-mat/9901019 (1999). 
\bibitem{exper1}
V. M. Pudalov, JETP Lett. {\bf 66}, 175 (1997); D. Popovic et al., 
Phys. Rev. Lett. {\bf 79}, 1543 (1997); N. Kim et al., cond-mat/9809357;
 D. Simonian, et al., Phys. Rev. Lett. {\bf 79},
2304 (1997); S. V. Kravchenko et al., Phys. Rev. B {\bf 58}, 3535 (1998);
S. V. Kravchenko et al., cond-mat/9812389 (1998). 
\bibitem{exper2} P. M. Coleridge et al., Phys. Rev. B {\bf 56}, R12764 (1997);
Y. Hanein et al., Phys. Rev. Lett. {\bf 80} 1288 (1998); 
M. Y. Simons et al., {\it ibid.} {\bf 80} 1292 (1998); 
S. J. Papadakis and M. Shayegan, Phys. Rev. B {\bf 57} R15068 (1998); 
Y. Hanein et al., {\it ibid.}  {\bf 58} R13338 (1998). 
\bibitem{class} F. Wegner, Nucl. Phys. B {\bf 36}, 663 (1989);
S. Hikami, Prog. Theor. Phys. Suppl. {\bf 107}, 213 (1992).
\bibitem{three} M. Batsch et al., Phys. Rev. Lett. 
{\bf 77}, 1552 ( 1996); K. Slevin and T. Ohtsuki, {\it ibid.} 
{\bf 78}, 4083 ( 1997); T. Drose et al., Phys. Rev. B
{\bf 57}, 37 (1998). 
\bibitem{qhe} B. Huckestein and B. Kramer,  Phys. Rev. Lett. {\bf 64}, 
1437 (1990); B. Huckestein, Europhys. Lett. {\bf 20}, 451 (1992); 
Y. Huo et al., Phys. Rev. Lett. ${\bf
70}$, 481 (1993).
\bibitem{super} Y. Liu et al., Phys. Rev. Lett. {\bf 67}, 
2068(1991).
\bibitem{fisher} M. P. A. Fisher, Phys. Rev. Lett. {\bf 65}, 
923(1990); M. P. A. Fisher  et al., 
{\it ibid.} {\bf 64}, 587 (1990).
\bibitem{anderson} P. W. Anderson, Phys. Rev. {\bf 109}, 
1492(1958).
\bibitem{lan} D. S. Fisher and P. A. Lee, Phys. Rev. B
${\bf 23}$, 6851 (1981); H. U. Baranger and A. D. Stone, {\it ibid.}
${\bf 40}$, 8169 (1989).
\bibitem{dns1} D. N. Sheng and Z. Y. Weng,  Phys. Rev. B {\bf 59}, 
(1999).
\bibitem{shahar}   D. Shahar et al., Science ${\bf 274}$,  589 (1996).

\end{references}
\end{document}